\documentclass[prb, 11pt,superscriptaddress,floatfix,nofootinbib,notitlepage,longbibliography]{revtex4-1}
\usepackage[utf8]{inputenc}
\usepackage{graphicx}
\usepackage[colorlinks=true,citecolor=blue]{hyperref}
\usepackage{amsmath}
\usepackage{amssymb}
\usepackage{soul}
\usepackage{gensymb}
\usepackage{amsfonts,amssymb,amsbsy}
\usepackage{siunitx}
\usepackage{float}
\usepackage{physics}
\usepackage{booktabs}
\usepackage[percent]{overpic}

\newcommand{\dIdV}{d$I$/d$V$~}
\usepackage{xcolor}
\usepackage{natbib}

\usepackage{lineno}

\begin{document}

	\title{Atomic-scale visualization of multiferroicity in monolayer NiI$_2$}
	
	\author{Mohammad Amini}
	\thanks{These two authors contributed equally}
	\affiliation{Department of Applied Physics, Aalto University, FI-00076 Aalto, Finland}
	
	\author{Adolfo O. Fumega}
	\thanks{These two authors contributed equally}
	\affiliation{Department of Applied Physics, Aalto University, FI-00076 Aalto, Finland}
	
	\author{H\'ector Gonz\'alez-Herrero}
	\affiliation{Department of Applied Physics, Aalto University, FI-00076 Aalto, Finland}
	\affiliation{Departamento de F\'isica de la Materia Condensada, 
		Universidad Aut\'onoma de Madrid, E-28049 Madrid, Spain}
	\affiliation{Condensed Matter Physics Center (IFIMAC), Universidad Aut\'onoma de Madrid, E-28049 Madrid, Spain}
	
	\author{Viliam Va\v{n}o}
	\affiliation{Department of Applied Physics, Aalto University, FI-00076 Aalto, Finland}
	\affiliation{Joseph Henry Laboratories and Department of Physics, Princeton University, Princeton, NJ 08544, USA}
	
	\author{Shawulienu Kezilebieke}
	\affiliation{Department of Physics, Department of Chemistry and Nanoscience Center, 
		University of Jyväskyl\"a, FI-40014 University of Jyväskyl\"a, Finland}
	
	\author{Jose L. Lado}
	\affiliation{Department of Applied Physics, Aalto University, FI-00076 Aalto, Finland}
	
	\author{Peter Liljeroth}
	\affiliation{Department of Applied Physics, Aalto University, FI-00076 Aalto, Finland}
	
	\email{}
	
	\begin{abstract}
		Progress in layered van der Waals materials has resulted in the discovery of ferromagnetic and ferroelectric materials down to the monolayer limit.\cite{Huang2017,Yuan2019} Recently, evidence of the first purely two-dimensional multiferroic material was reported in monolayer NiI$_2$.\cite{Song2022} However, probing multiferroicity with scattering-based and optical bulk techniques is challenging on 2D materials,\cite{Jiang2023} and experiments on the atomic scale are needed to fully characterize the multiferroic order at the monolayer limit. Here, we use scanning tunneling microscopy (STM) supported by theoretical calculations based on density functional theory (DFT) to probe and characterize the multiferroic order in monolayer NiI$_2$. We demonstrate that the type-II multiferroic order displayed by NiI$_2$, arising from the combination of a magnetic spin spiral order and a strong spin-orbit coupling,\cite{Fumega2022} allows probing the multiferroic order in the STM experiments. Moreover, we directly probe the magnetoelectric coupling of NiI$_2$ by external electric field manipulation of the multiferroic domains. Our findings establish a novel point of view to analyse magnetoelectric effects at the microscopic level, paving the way towards engineering new multiferroic orders in van der Waals materials and their heterostructures.

	\end{abstract}
	
	\date{\today}

	\maketitle

	\section*{Introduction}
	
	Multiferroics are materials that exhibit simultaneously more than one ferroic order.\cite{Hill2000,Fiebig2016,Spaldin2019} 
	Over the last years, numerous studies have reported different bulk multiferroics, most of them complex oxides.\cite{Kimura2003,Hur2004,Gajek2007,Nan2008} 
	Multiferroics can be classified based on the presence of weak or strong coupling between the orders as type I or type II, respectively.\cite{Fiebig2016,Spaldin2019}
	In the particular case of magnetic  and ferroelectric orders, type-II multiferroics showing a strong magnetoelectric coupling\cite{Fiebig2005,doi:10.1126/science.1260561} have a huge potential for technological applications including spintronics, data storage, and efficient energy management in computation.\cite{Pantel2012,Hu2015,Schoenherr2020} Nevertheless, multiferroics displaying a strong magnetoelectric coupling at sufficiently high temperatures to build functional devices remain elusive. 
	
	The dawn of two-dimensional (2D) materials has enabled new strategies for the design of artificial multiferroics.\cite{vdwHT2013} 
	The weak van der Waals bonding between the layers allows to easily reach the 2D limit with these compounds, thus obtaining families of building blocks with different properties.\cite{Fei2018,Huang2017,doi:10.1021/acs.nanolett.6b03052,Gong2017,doi:10.1021/acs.nanolett.9b00553,NbSe22015,Ising2018,doi:10.1021/acs.nanolett.7b04852, Yuan2019} These can be stacked and twisted in heterostructures, leading to a wide range of emergent phenomena including multiferroicity.\cite{mottCao2018,superCao2018,Kezilebieke2020,Vano2021,10.21468/SciPostPhys.13.3.052,Fumega_2023} 
	Recently, it was shown that the multiferroicity of the bulk van der Waals compound NiI$_2$ remains in the mechanically-exfoliated few-layer\cite{Ju2021} and monolayer limits\cite{Song2022}. Theoretical analyses have addressed the origin of this type-II multiferroic order to the combination of the magnetic spin-spiral order of NiI$_2$ and the strong spin-orbit (SOC) coupling of the I atoms.\cite{Fumega2022}
	To provide experimental evidence of 2D multiferroicity, circular dichroic Raman, birefringence, and second-harmonic-generation (SHG) measurements on both mono- and multi-layer NiI$_2$ were performed, establishing NiI$_2$ as the first purely 2D multiferroic with a transition temperature $T_C=21$ K.\cite{Song2022}
	However, optical measurements might not suffice to demonstrate the emergent ferroelectricity in the 2D limit.\cite{Jiang2023} In addition, typical bulk techniques, such as neutron scattering, used to identify the magnetic spin spiral that gives rise to the multiferroic order, cannot be applied in the monolayer limit.\cite{BILLEREY1977138} 
	These factors suggest that the origin of the multiferroic order in monolayer NiI$_2$ has not been experimentally established and its full characterization requires measurements at the microscopic level.

	In this work, we present real-space visualization of the multiferroicity in monolayer NiI$_2$. We show that the magnetoelectric coupling present in this kind of type-II multiferroics allows us to directly probe and characterize the multiferroic order by scanning tunneling microscopy (STM). These results are supported by non-collinear \emph{ab initio} calculations based on density functional theory (DFT) in monolayer NiI$_2$. Moreover, we provide evidence of the magnetoelectric coupling of this 2D system by electrical manipulation of the multiferroic domains. Our findings demonstrate the visualization of multiferroic order in van der Waals materials with
	atomic-scale resolution, establishing a strategy that can be further applied to artificial multiferroic van der Waals heterostructures.

	\begin{figure}[t!]
		\centering
		\includegraphics[width = \textwidth]{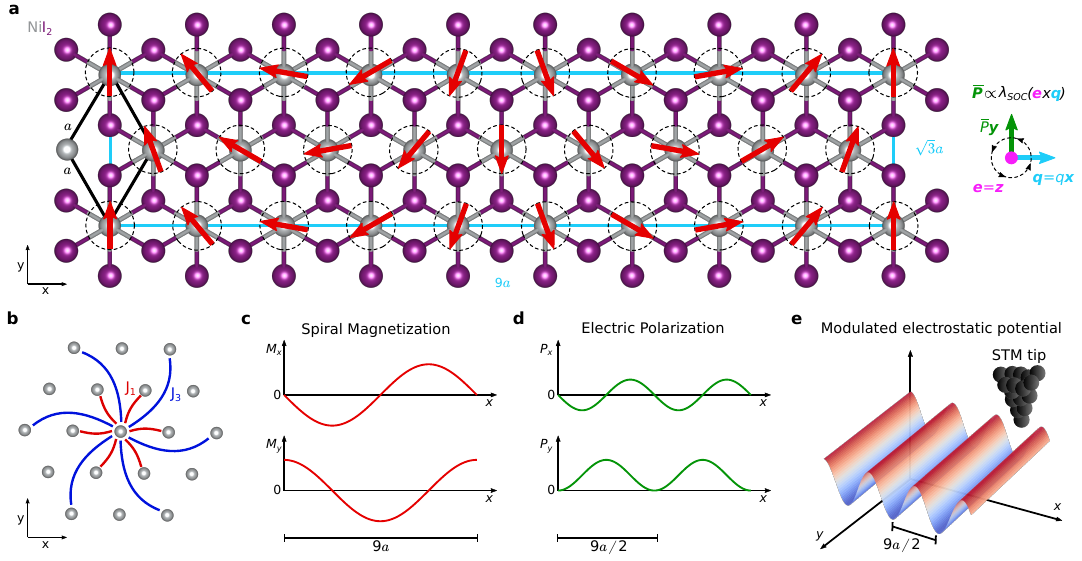}
		\caption{\textbf{a:} Unit cell of monolayer NiI$_2$. $9a\times\sqrt{3}a$ supercell is required to describe the magnetic spin-spiral order with propagation vector $\mathbf{q}$ in the $x$-direction and rotation vector $\mathbf{e}$ in the z-direction. The presence of spin-orbit coupling $\lambda_\mathrm{SOC}$ induces a net electric polarization $\overline{P}$ in the y-direction. \textbf{b:} Ferromagnetic first neighbor $J_1$ and antiferromagnetic third neighbor $J_3$ magnetic exchange interactions between Ni atoms giving rise to the spin spiral order in the monolayer. \textbf{c:} Magnetization components of the spin spiral showing a periodicity of $9a$ in the $x$-direction. \textbf{d:} Electric polarization components associated with the magnetic spin spiral in the presence of  SOC showing a periodicity of $9a/2$ in the $x$-direction. \textbf{e:} Modulated electrostatic potential that can be probed with an STM. It shows a $9a/2$ periodicity in the $x$-direction as a consequence of the multiferroic order.}
		\label{Figure1}
	\end{figure}

	\section*{Multiferroic order in N\lowercase{i}I$_2$}
	
	In bulk, NiI$_2$ displays a magnetic spin spiral at low temperatures.\cite{BILLEREY1977138} 
	DFT calculations have determined that the $\mathbf{q}$ vector characterizing the spin spiral order in bulk is a consequence of the competition between magnetic exchange interactions between Ni atoms\cite{PhysRevB.106.035156,Fumega2022,2023arXiv230604729K}. In particular, ferromagnetic intralayer first-neighbor $J_1$, antiferromagnetic intralayer third-neighbor $J_3$, and antiferromagnetic interlayer second-neighbor $\tilde{J}_2$ magnetic exchange interactions are the most relevant.
	These interactions lead to a $\mathbf{q}$ vector in bulk whose in-plane component lies on the $\overline{\Gamma M}$ segment in reciprocal space, and whose value can be modified by external parameters, such as pressure, that change the ratio between the different magnetic exchanges.\cite{PhysRevB.106.035156,2023arXiv230611720O} In the monolayer limit, there are no interlayer interactions, hence the $\mathbf{q}$ vector is determined by the competition between $J_1$ and $J_3$ (Fig.~\ref{Figure1}b). DFT calculations have predicted that in this scenario the $\mathbf{q}$ vector lies on the $\overline{\Gamma K}$ segment in reciprocal space, corresponding to the third-neighbor bond direction.\cite{Sodequist_2023} The classical Heisenberg model solution predicts that $\mathbf{q}=(q,q,0)$, with $q$ in units of the reciprocal lattice of the structural single-unit cell with lattice vector $\mathbf{a}$, is determined by the $J_1/J_3$ ratio.\cite{PhysRevB.93.184413,PhysRevB.104.184427}

	Apart from the magnetic order, theoretical analyses have shown that monolayer NiI$_2$ develops a ferroelectric polarization due to the combination of this spin spiral order and the strong 
	spin-orbit coupling of the I atoms.\cite{Fumega2022} These results are based  on the theoretical derivation for the emergence of ferroelectricity in spiral magnets.\cite{PhysRevLett.96.067601} This approach predicts that a net ferroelectric polarization $\overline{\mathbf{P}}$, which is proportional to the SOC strength $\lambda_\mathrm{SOC}$, arises perpendicular to the $\mathbf{q}$ and $\mathbf{e}$ rotation vector that characterizes the spin spiral (as shown in Fig.~\ref{Figure1}a). At the more microscopic level, this relationship shows that given a spin spiral with magnetization $\mathbf{M}=\mathrm{M}(-\sin(\mathbf{q}\mathbf{r}), \cos(\mathbf{q}\mathbf{r}),0)$ in cartesian coordinates (Fig. \ref{Figure1}c) the emergent electric polarization is given by
	
	\begin{equation}\label{eq:P_spiral}
		\mathbf{P}=\Lambda\frac{\mathbf{M}\times(\nabla \times \mathbf{M})}{\mathrm{M}^2},
	\end{equation}
	
	where $\Lambda$ is proportional to $\lambda_\mathrm{SOC}$ and some physical constants.\cite{PhysRevLett.100.077202} 
	Equation (\ref{eq:P_spiral}) leads to $\mathbf{P}=\Lambda \mathrm{q}(-\sin(2\mathbf{q}\mathbf{r}), \sin^2(\mathbf{q}\mathbf{r}),0)$ for the given spin spiral $\mathbf{M}$.
	The spatial average of this equation leads to a net polarization $\overline{\mathbf{P}}$ with non-zero components only in the direction perpendicular to the $\mathbf{q}$ and $\mathbf{e}$ vectors.\cite{PhysRevLett.96.067601, PhysRevLett.100.077202}
	Equation (\ref{eq:P_spiral}) shows that the emergent electric polarization has a real space modulation whose periodicity is half of the periodicity of the spin spiral (Fig.~\ref{Figure1}d). 
	
	The modulation of the induced ferroelectric order, which is produced by the coupling between spin and charge degrees of freedom, is the property that can be used to characterize the multiferroic order of monolayer NiI$_2$ with an STM.
	The emergence of ferroelectricity is accompanied by ferroelectric displacements of the atomic positions in the directions of the electric polarizations and results in the emergence of a modulated electrostatic potential. 
	This electrostatic potential displays the same periodicity as the ferroelectric polarization and modulates the energy bands of NiI$_2$. The modulation results in an observable contrast in STM images, thus providing a signature of the multiferroic order allowing its characterization (Fig.~\ref{Figure1}e). 
	The magnetoelectric coupling occurring in this multiferroic allows the visualization of the magnetic spin spiral periodicity without the need to use more complex spin-polarized scanning microscopy techniques.\cite{doi:10.1126/science.abd9225,Schoenherr2018}

	\section*{STM characterization of monolayer N\lowercase{i}I$_2$}
	
	\begin{figure*}[t!]
		\centering
		\includegraphics[width = 1\textwidth]{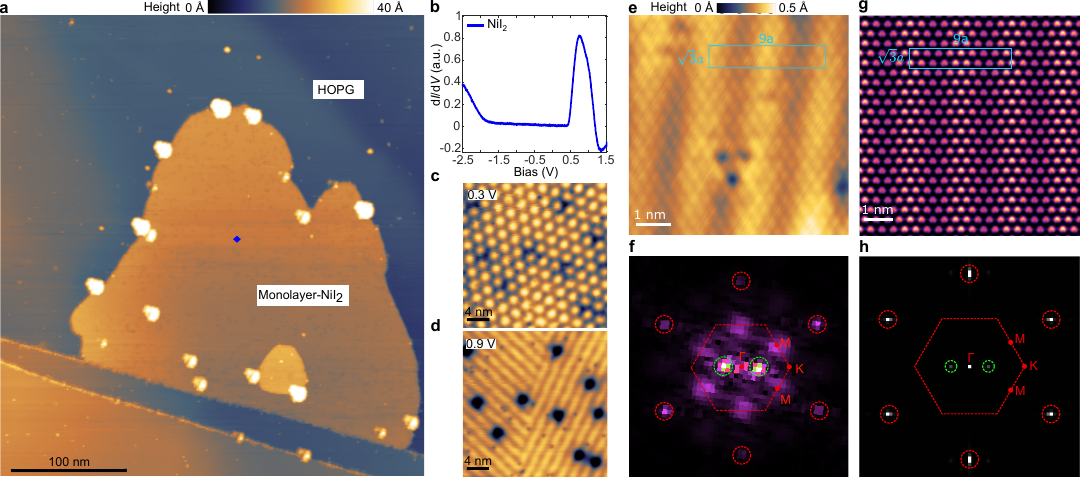}
		\caption{\textbf{a:} Large area STM scan of monolayer NiI$_2$ on HOPG  (image size:  $ 370\times 400 \mathrm{~nm}^{2}, V= 1.2 \mathrm{~V}, I= 4 \mathrm{~pA} $). \textbf{b:} d$I$/d$V$ point spectra measured on monolayer NiI$_2$ (blue dot shown in panel a). \textbf{c-d:} Small-area STM images of monolayer NiI$_2$ at $0.3 \mathrm{~V}$ and $0.9 \mathrm{~V}$, respectively (image size:  $ 25\times 25 \mathrm{~nm}^{2} $). \textbf{e-f:} Atomic resolution STM scan of monolayer NiI$_2$ (image size  $ 6.5\times 6.5 \mathrm{~nm}^{2}, V= 0.43 \mathrm{~V}, I= 100 \mathrm{~pA} $) and corresponding FFT. \textbf{g-h:} DFT-computed STM image at the conduction band of monolayer NiI$_2$ and corresponding FFT. In the real space images, the light blue rectangle depicts the unit cell of the commensurate spin spiral shown in Fig.~\ref{Figure1}a. In contrast to the theory, the experimental stripes are not aligned precisely with the atomic lattice. In the FFT images, the red-circled peaks correspond to the single unit cell giving rise to the depicted Brillouin zone, while the green-circled peaks are associated with the ferroelectric modulation.}
		\label{Figure2}
	\end{figure*}
	
	Figure \ref{Figure2}a shows a large area STM scan of our sample (see SI for details). It reveals the presence of a well-formed monolayer NiI$_2$ on HOPG. 
	The \dIdV spectrum of monolayer NiI$_2$ is shown in Fig.~\ref{Figure2}b. The valence and conduction bands can be distinguished at $-1.9$ V and $0.4$ V respectively, indicating that NiI$_2$ is an insulator with a gap of $2.3$ eV.
	Figures \ref{Figure2}c-d show small-area STM images at $0.3$ V (within the NiI$_2$ bandgap) and at $0.9$ V (within the NiI$_2$ conduction band). A systematic analysis with scans at different energies can be found in the SI.
	Inside the gap, the contrast is not affected by the polarization of the NiI$_2$, and we observe a hexagonal moir\'e modulation arising from the lattice mismatch between HOPG and NiI$_2$ (Figure \ref{Figure2}c). However, as the conduction band is modulated by the ferroelectric polarization, the stripe-like pattern expected for this kind of type-II multiferroic appears at biases corresponding to the NiI$_2$ conduction band (Fig.~\ref{Figure2}d). Two distinct regions with different stripe orientations can be observed. These correspond to two different multiferroic domains. Finally, there are different types of atomic scale defects that are analysed in detail in the SI.

	Figure \ref{Figure2}e shows an atomic-resolution scan in a single domain region, the FFT of this scan is shown in Fig.~\ref{Figure2}f. The spin-spiral $\mathbf{q}$ vector can be determined from the relationship between the single unit cell $a\times a$ and the stripe modulation. 
	From our experimental results, we obtain $a=3.85$ Å for monolayer NiI$_2$. The periodicity of the stripes is $L_S=17.8$ Å corresponding to about $4.6a$.
	According to Eq.~(\ref{eq:P_spiral}) the modulation of the stripes caused by the emergent electric polarization displays at half the periodicity of the spin spiral (Figs.~\ref{Figure1}c-d), leading to a spin spiral periodicity of $9.2a$ in the experiment. The spin-spiral $\mathbf{q}$ vector can be directly extracted from the peaks associated with the stripes in the FFT (green circles in Fig.~\ref{Figure2}f) as half of the peak value. We have obtained $\mathbf{q}=(0.069,0.041,0)$ in units of the reciprocal lattice vectors, showing a small deviation from $\overline{\Gamma K}$ segment expected for the $J_1-J_3$ spin model. The projection in the $\overline{\Gamma K}$ segment is $0.173\overline{\Gamma K}$, corresponding to $\mathbf{q}=(0.057,0.057,0)$. It can be observed that this small variation in the experimental $\mathbf{q}$ vector from the $\overline{\Gamma K}$ segment leads to a small tilt ($\sim4.6\degree$) of the stripe modulation (Fig.~\ref{Figure2}f). This effect could be a consequence of local defects in the sample, the proximity of other multiferroic domains, or the influence of other parameters such as intralayer second-neighbor magnetic exchange, Kitaev, or biquadratic interactions not considered in the $J_1-J_3$ spin model.\cite{PhysRevLett.131.036701,PhysRevB.104.184427} Finally, the diffuse peaks around the M points are associated to the defect seen in the scan. The analysis of the filtered FFT images is shown in the SI.

	Extracting the information on the spin spiral from the STM measurements allows us to model monolayer NiI$_2$ using DFT. Neglecting the small tilt in the ferroelectric modulation, the unit cell describing the multiferroic order in monolayer NiI$_2$ can be well approximated by the commensurate $9a\times\sqrt{3}a$ supercell shown in Fig.~\ref{Figure1}a,
	corresponding to $\mathbf{q}=(0.064,0.064,0)$ and $\sqrt{3}/9\overline{\Gamma K}\sim0.192\overline{\Gamma K}$.
	From the experimental $\mathbf{q}=(0.057,0.057,0)$ we can estimate $J_3/J_1=-0.263$.\cite{PhysRevB.93.184413,PhysRevB.104.184427}  
	This result shows that monolayer NiI$_2$ is on the verge of a spin-spiral to ferromagnetic transition ($J_3/J_1=-0.25$),\cite{PhysRevB.104.184427} which might explain why in some studies the presence of ferroelectric polarization was only reported down to the bilayer limit but not in the monolayer.\cite{Ju2021} 
	It also points out that DFT tends to overestimate the $|J_3/J_1|$ ratio.\cite{Sodequist_2023,2023arXiv230611720O} This can be rationalized by the typical limitations of the different DFT functionals to deal with electronic correlations and electronic localization, thus introducing an overestimation of the long-range magnetic exchanges. This could explain why \emph{ab initio} calculations predict a spin spiral for NiCl$_2$ while the observed ground state is ferromagnetic.\cite{Sodequist_2023}
	
	\begin{figure}[t!]
		\centering
		\includegraphics[width = 1\textwidth]{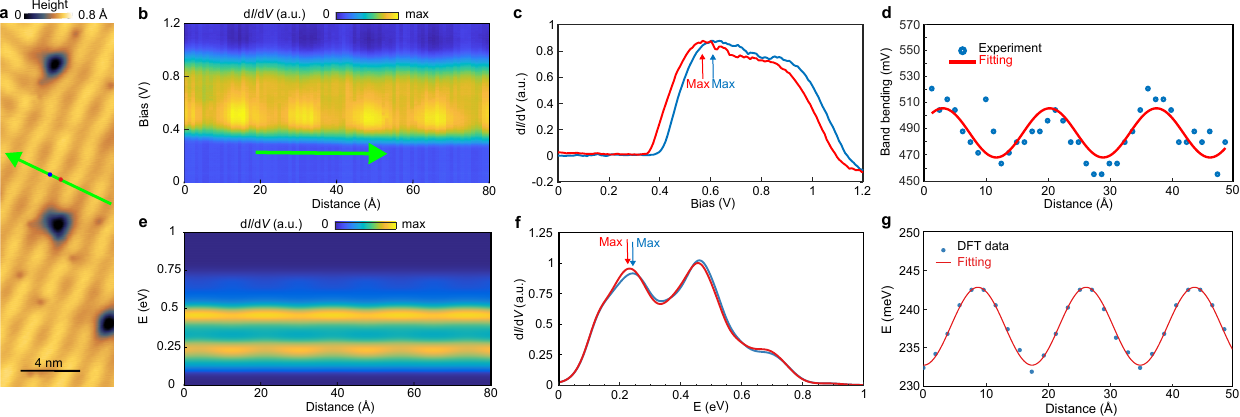}
		\caption{\textbf{a:} Small area STM scan showing a multiferroic domain where a line spectrum was performed along the green arrow ($V= 1 \mathrm{~V}, I= 100 \mathrm{~pA} $). \textbf{b:} d$I$/d$V$ line spectrum capturing the conduction band of monolayer NiI$_2$ along the green arrow in panel (a) parallel to the $\mathbf{q}$ vector.\textbf{c:} d$I$/d$V$ spectra showing the shift in energy between a maximum and a minimum of the stripe modulation (blue and red dots in panel (a))\textbf{d:} d$I$/d$V$ intensity maximum of the bottom of the conduction band along the green line showing a modulated band bending as a consequence of the ferroelectric modulation. \textbf{e-g:} Corresponding DFT calculations for panels (b-d) computed in the $9a\times\sqrt{3}a$ supercell and the $\mathbf{q}=(0.064,0.064,0)$ spin spiral.
		}
		\label{Figure3}
	\end{figure}
	
	In order to shed light on the microscopic origin of the STM modulation observed in the experiments, we have performed non-collinear DFT calculations in the commensurate $9a\times\sqrt{3}a$ supercell and the $\mathbf{q}=(0.064,0.064,0)$ spin spiral (see SI for computational details). 
	When SOC is included in the atomic relaxations, we observe the emergence of ferroelectric forces that drive the atoms to new positions following the same periodicity as the electric polarization (Fig. \ref{Figure1}d). Due to this symmetry breaking in real space, the computed STM image (Fig.~\ref{Figure2}g) and its FFT (Fig.~\ref{Figure2}h) display a modulation with the periodicity of the electric polarization, i.e. half of the magnetic spiral, in a good agreement with the experimental observations.  
	
	Having characterized the spiral order of monolayer NiI$_2$, we now proceed to determine the strength of the ferroelectric polarization. Local variations of the electric polarization, such as those encountered on the edges of ferroelectric islands, induce a local energy band bending that can be used to estimate the strength of the electric polarization.\cite{Chang2016, Chang2020,Amini2023}  
	The emergence of an inhomogeneous electrical polarization in NiI$_2$ is expected to produce similar band bendings, but in this case within the same multiferroic domain and with the periodicity of the ferroelectric modulation (Fig.~\ref{Figure1}d). This behavior is experimentally demonstrated and theoretically reproduced in Fig.~\ref{Figure3}.
	
	Figure \ref{Figure3}b shows \dIdV spectra recorded along a line in a single domain region (green arrow shown in Fig.~\ref{Figure3}a) parallel to the $\mathbf{q}$ vector. Figure \ref{Figure3}d shows the DFT simulated line spectra for the commensurate $9a\times\sqrt{3}a$ supercell and the $\mathbf{q}=(0.064,0.064,0)$ spin spiral. Both plots show the conduction bands. A clear underestimation of the band gap is observed for the DFT calculations, which is typical for the LDA approximation. However, both experiment and theory show a band bending modulation of the conduction band that becomes more apparent when analyzing the \dIdV spectra at a maximum and a minimum of the stripe modulation in Fig.~\ref{Figure3}c (spectra taken at the positions indicated by the blue and red dots in Fig.~\ref{Figure3}a, respectively). A clear shift in energies between these two positions occurs as a consequence of the ferroelectric modulation introducing a band bending. The shift in energies is also seen in the DFT calculations (Fig.~\ref{Figure3}f). The band bending can be analyzed by tracking the shift in the differential conductance maxima as a function of the position across the modulation. This is shown in Figs.~\ref{Figure3}d and  \ref{Figure3}g for the experimental and theoretical data, respectively. The observed oscillatory behavior can be fitted to $E=E_0+E_P\sin(2\pi x/L_S+\phi)$, where $E_0$ corresponds to an average of the differential conductance maximum at the bottom of the conduction band, $E_P$ is the band bending modulation caused by the inhomogeneous emergent polarization, $L_S$ is the stripe periodicity that has already been analyzed and, $\phi$ is a trivial phase related to the starting point considered for the fitting. We obtain $E_P=16.8$ mV from the experiment and $E_P=5.0$ meV from the DFT calculation. This underestimation of the \emph{ab initio} calculations could be a consequence of the limitations of the LDA approximation for the atomic relaxations. However, apart from this small difference, the qualitative agreement between theory and experiment is remarkable. For phonon-driven ferroelectrics such as SnSe and SnTe, STM experiments have shown the band bending is on the order of several hundreds of meV\cite{Chang2016,https://doi.org/10.1002/adma.201804428, Chang2020,Amini2023} and an associated electric polarization $P\sim 10^{–10}$ C/m has been estimated.\cite{Chang2020,https://doi.org/10.1002/adma.201804428} Therefore, considering that in monolayer NiI$_2$ we observe a band bending of 1 or 2 orders of magnitude smaller, a $P\sim 10^{–12}$ C/m can be estimated for monolayer NiI$_2$, consistent with previous theoretical estimations.\cite{Fumega2022}
	
	\section*{Manipulation of multiferroic domain boundaries}
	
\begin{figure}[t!]
		\centering
		\includegraphics[width = 1\textwidth]{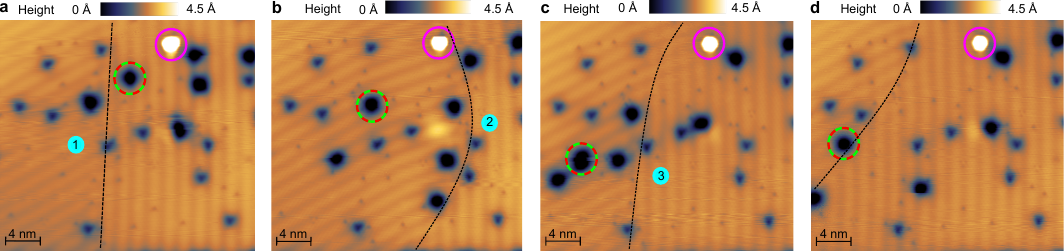}
		\caption{\textbf{a-d:} Manipulation of the multiferroic domains by sweeping the bias from 1 V to 4 V with closed feedback loop (image size  $ 30\times 27 \mathrm{~nm}^{2}, V= 1 \mathrm{~V}, I= 20 \mathrm{~pA} $). Sky blue dots mark the position of the tip during the process and the numbers on them track the order of events. The black-dashed lines highlight the domain boundary, the pink circle shows a neutral defect on the surface used as a spacial reference, and the dashed red-and-green circle corresponds to a type of mobile charge defect.}
		\label{Figure4}
	\end{figure}

	We have also directly probed the magnetoelectric coupling in monolayer NiI$_2$  by external electric field manipulation of the multiferroic domains.
	Figs.~\ref{Figure4}a-\ref{Figure4}d show the movement of the multiferroic domain wall induced by voltage pulses from the STM tip (the bias is swept from 1 V to 4 V with a closed feedback loop and a current set point of 100 pA). Starting in Fig.~\ref{Figure4}a, the multiferroic domain wall (black-dashed line) and different kinds of defects (highlighted with circles, detailed analysis in the SI) can be identified. The neutral defects (pink circle) can be used as a spatial reference for the domain manipulation.
	A first voltage pulse is performed in the position indicated by the sky blue dot (in the left domain) and it leads to the configuration shown in Fig.~\ref{Figure4}b. We can observe that the domain wall has moved to the right increasing the size of the left domain.  
	Moreover, a clear difference between the different types of defects can be observed. While the charged defects are mobile the neutral ones do not move and can be used as a reference to analyze the domain manipulation. After the scan shown in Fig.~\ref{Figure4}b, we applied a second voltage pulse in the right domain leading to Fig.~\ref{Figure4}c, and a third one leading to Fig.~\ref{Figure4}d. As before, the location of the domain wall moves away from the position of the voltage pulse and charged defects are mobile, while neutral ones are not.
	Ferroelectric domain wall manipulation through voltage pulses had been achieved before in ferroelectric SnTe.\cite{Amini2023} Now, we show here that multiferroic domains in this kind of magnetic spin-spiral multiferroics are also tunable by external electric fields, thus showing direct evidence of the magnetoelectric coupling in monolayer NiI$_2$.

	\section*{Conclusions}
	In conclusion, we have probed and characterized the multiferroic order in monolayer NiI$_2$ down to the atomic scale. Using a combination of STM experiments and non-collinear \emph{ab initio} calculations, we show that in this spin-spiral SOC-driven multiferroic, the emergent ferroelectric polarization induces a modulation of the electrostatic potential. This is directly reflected in the local density of states allowing its direct visualization by STM. This modulation has half of the spin-spiral periodicity, allowing us to characterize the spin-spiral vector as $\mathbf{q}=(0.069,0.041,0)$ and the parameter relation $J_3/J_1=-0.263$ of the associated spin model. The observed band bending allows us to estimate the polarization $P\sim 10^{–12}$ C/m for monolayer NiI$_2$. Finally, we have probed the magnetoelectric coupling of NiI$_2$ by manipulating the multiferroic domains by local electric fields induced by the STM tip. Our results firmly establish the atomic scale origin of multiferroicity in NiI$_2$ and pave the way for future studies on multiferroics and 2D van der Waals materials at the microscopic level.

	\section*{Acknowledgements}
	This research made use of the Aalto Nanomicroscopy Center (Aalto NMC) facilities and was supported by the European Research Council (ERC-2017-AdG no.~788185 ``Artificial Designer Materials'' and ERC-2021-StG no.~101039500 ``Tailoring Quantum Matter on the Flatland'') and Academy of Finland (Academy professor funding nos.~318995 and 320555, Academy research fellow nos.~331342, 336243 and nos.~338478 and 346654, and Academy postdoctoral fellow no. 349696). Computing resources from the Aalto Science-IT project and CSC Helsinki are gratefully acknowledged. V.V. acknowledges fellowship support from the Princeton Center for Complex Materials supported by NSF-DMR-2011750. 

	\bibliography{NiI2}

\end{document}